# Examining The CoVCues Dataset: Supporting COVID Infodemic Research Through A Novel User Assessment Study

Shreetika Poudel
*School Of Computing and Analytics*
*Northern Kentucky University*
Highland Heights, Kentucky, USA
poudels2@nku.edu

Ankur Chattopadhyay, PhD
*School Of Computing and Analytics*
*Northern Kentucky University*
Highland Heights, Kentucky, USA
chattopada1@nku.edu

***Abstract* - The public confidence and trust in online healthcare information have been greatly dented following the COVID-19 pandemic, which triggered a significant rise in online health misinformation. The World Health Organization (WHO) has called this problem the COVID infodemic. Existing literature shows that different datasets have been created to aid with detecting false information associated with this COVID infodemic. However, most of these datasets contain mostly unimodal data, which comprise primarily textual cues, and not visual cues, like images, infographics, and other graphic data components. Prior works point to the fact that there are only a handful of multimodal datasets that support COVID misinformation identification, and they lack an organized, processed and analyzed repository of visual cues. The novel CoVCues dataset, which represents a varied set of image artifacts, addresses this gap and advocates for the use of visual cues towards detecting online health misinformation. CoVCues is a uniquely exclusive collection of images for which we make use of categorization and sub-categorization of images from publicly available data. We also utilize AI models to show the capability of visual cues in enhancing misinformation detection. As part of validating the contents and utility of our CoVCues dataset, we have conducted a preliminary user assessment study, where different participants have been surveyed through a set of questionnaires to determine how effectively these dataset images contribute to the user perceived information reliability. The survey participants were shown sample images from this dataset and were asked to differentiate between reliable and unreliable content. These survey responses helped provide early insights into how different stakeholder groups interpret visual cues in the context of online health information and communication. The findings from this novel user assessment study offer valuable feedback for refining our CoVCues dataset and for supporting our claim that visual cues are underutilized but useful in combating the COVID infodemic. Overall, our CoVCues dataset paves the path towards a new image-centric approach to empower COVID misinformation detection and presents a valuable resource for both researchers and professionals in fighting against the COVID infodemic. To our knowledge, this user assessment research study, as described in this paper, is the first of its kind work, involving COVID visual cues, that demonstrates the important role that our CoVCues dataset can potentially play in aiding COVID infodemic related future research work.**

## I. Introduction And Motivation

The COVID-19 pandemic [1, 2, 16] has witnessed a widespread growth in public health related misinformation and fake news with the COVID infodemic issue emerging as an important topic of discussion. The spread of health misinformation in relation to the COVID virus has posed as a major threat to public health and international relationships. From dangerous health advice, like consuming bleach for curing the infection to politically motivated conspiracy theories about the virus origins, misinformation can seriously impact the viewers' health and perception. The rampant spread of false and misleading health information, its vectors of transmission, available treatments, need to be addressed through active and more extensive research [8]. For instance, the gravity of this infodemic issue is manifested by the fact that conspiracy theories falsely linking 5G telecommunications infrastructure with the exacerbation of COVID symptoms have led people to burn down mobile phone towers, placing both human life and property in danger [5].

An ensuing lack of faith in public healthcare communications and a reduction of trust in online healthcare information [12] has led to a global crisis, which has gained prominence as a critical area of research since 2020. Even now as we have entered the post COVID pandemic period, health misinformation continues to be a relevant research problem with fabricated articles and manipulated contents posted online still deceiving end-users till date. This has resulted in reliability concerns in online content consumers, including information seekers and patients, thus impacting their trust in content providers, including the healthcare institutions [16]. Furthermore, this has created hesitation to vaccinate against diseases, such as COVID. This entire infodemic phenomenon has negatively affected public health and safety.

After the COVID pandemic, various datasets [2] have been published in an effort to assist with misinformation detection and to study this infodemic problem. However, these datasets have mostly prioritized textual cues and lack a detailed study of visual cues, especially a rigorous analysis of image artifacts. Given the recent emphasis on utilizing visual cues in misinformation detection, we have developed the new CoVCues dataset [15], which is an exclusive repository of a unique collection of image artifacts and makes use of categories plus sub-categories to effectively organize and classify these visual cues. As part of a maiden research study, we perform a substantial analysis of these dataset images using AI models to study them, as discussed in this paper. We also conduct a first-of-its-kind user study, as reported in this paper, to examine how these images can be used to increase information assurance for users viewing online health information (OHI) [12].

In order to tackle the infodemic effects, previous research & development efforts have seen the emergence of datasets, like MM-Covid, CoAID, MMCoVaR, and ReCOVery [2, 11], in attempts to boost COVID misinformation detection by incorporating data from tweets, news articles, videos, and other social media contents. However, each of these datasets have specific limitations, and are lacking in context of multimodal data components, that include organized and categorized image artifacts. CoAID is a very large textual dataset and cannot be used for the detection of misinformation in multimedia contents. MM-Covid may cover multimodal cues, but it lacks a well-defined mechanism for structuring and labeling visual cues, including image features. Therefore, it is limited in terms of supporting misinformation detection through image pattern recognition. Similarly, even though ReCOVery encompasses a diverse set of multimedia contents, it is also hindered by insufficient organized taxonomy of visual cues, that diminishes the substantive application of images in the analysis of misinformation. These constraints highlight the necessity for a more extensive dataset that efficiently integrates and structures a variety of visual cues in the form of infographic indicators to improve the process of misinformation identification endeavors [6]. To address these shortcomings, there is a need to explore other methodologies, such as applying image contents, towards enhancing the accuracy and reliability of detecting misinformation, especially in health-related online posts [3].

The CoVcues dataset [15] was developed by compiling online sources using web URLs from pre-existing datasets and scraping them for images. To gain insight into the large amount of visual data cues, which were extracted from the COVID web contents, a classification taxonomy was used to sort images by their types. The built dataset was further analyzed using AI models to test its usefulness in misinformation detection and information assurance. To validate our approach and assess the real-world applicability of CoVCues, we conducted an initial user assessment study, where participants were asked to review sample images from our dataset and indicate which ones they perceived as reliable or unreliable. This allowed us to investigate how different stakeholders, including OHI consumers and OHI content providers [16], interpret visual cues and use them to make reliability judgements. We also gathered participants' perceptions of the importance of visual, textual, and combined cues when assessing the authenticity of OHI. Our study not only supports the utility of CoVCues, but also provides early evidence that images play a pivotal part in influencing user determination of credibility. These insights are critical for improving misinformation detection and for designing more effective, user-centered interventions. To the best of our knowledge, this is one of the first efforts to pair multimodal dataset development with a focus on visual cues and a user-centered assessment study, thus offering promising directions for future misinformation research around the COVID infodemic issue.

## II. Related Works

Since the internet's introduction, online users have been spreading misinformation, fake news, and conspiracy stories both intentionally and accidentally. As discussed earlier, this became a public concern during the COVID pandemic when WHO declared a public health emergency due to the infodemic, which has led to worse outcomes, including prolonged lockdowns, loss of public trust in vaccinations, doctors, service providers and the over politization of healthcare [4].

### *A. Visual Cues and Misinformation*

A previous work [10] examines how inattention contributes to the diffusion of misinformation on social media. Across two experiments comprising of a sample size of over 1,700 United States adults, this work found that users are more likely to share false information since a good number of people do not normally think about the accuracy of what they are going to share. The first experiment showed that participants were more correct in judging the truth versus falsehood of COVID headlines when explicitly asked about accuracy compared with when asked about willingness to share. This study also indicated that those higher in cognitive reflection and with more scientific knowledge tended to have a greater relative discernment of truth. This second experiment added only a very subtle manipulation: an "accuracy nudge," whereby participants were simply asked to rate an unrelated headline regarding accuracy before making sharing decisions. This resulted in an increased capacity of participants in differentiating true from false headlines, nearly tripling their level of truth assessment. These results suggest that accuracy prompts may become an efficient and practical means of reducing social media-based misinformation in the most critical crisis contexts, like the COVID pandemic. This approach is a genuinely practical way of raising the quality of information shared online without recourse to censorship of contents [9].

The findings from our unique study, as presented later in this paper, greatly pertain to researching the role of visual cues in detecting COVID misinformation. By providing room for reflection on how attention and cognitive reflection can be drawn to such processes, our study initiates a research space connected to how visual cues, such as warning labels or accuracy prompts, might guide users to critically assess the content encountered. Subtle infographic cues to accuracy embedded within social media could be that "nudge" for the user to take a second look and consider the truthfulness of information before sharing it. The premise of our work is further supported by evidence from an existing study that shows how minor interventions can create a huge impact on the quality of content shared online, hence making images a promising ally to counter misinformation [7].

## B. Existing COVID Datasets

Multiple unimodal datasets on COVID related research already exist. However, only a few notable ones with a potential scope of being expanded into multimodal cues, were identified as a starting point for forming our CoVCues dataset, and these include: CoAID, ReCoVery, MM-COVID, and MM-CoVaR, as discussed earlier. The online sources (URLs) from the first three of these mentioned datasets were used for our visual data cues collection, including the web scraping needed for building CoVCues. We strategized to build off of the existing source links in these datasets to ensure we are not further spreading online health misinformation. Each of these datasets has unique strengths and weaknesses for classifying and collecting reliable plus unreliable data components. However, due to the absence of an organized taxonomy for handling and categorizing visual cues, we acted upon the need to create a more sophisticated dataset, which could better integrate and explore multimodal data. Furthermore, these datasets do not adequately reflect the dynamic characteristics of misinformation, setting the stage for us to pursue the development of CoVCues.

CoAID is possibly the most well-known COVID misinformation dataset [2]. It was released in 2020 and contains 4,251 news articles, 296,000 related user engagements, and 926 social media posts. It classifies data as unreliable or reliable by using a third-party fact checking site such as “who.init” which is related to WHO. It is a single-modal textual dataset, and its extensive amount of source URLs made it a good choice to parse for images as part of building CoVCues. On the other hand, ReCOVery [2] is an extensive multi-modal dataset that was used to analyze the flow of misinformation and conspiracies from news articles to social media. It consists of 2,029 news articles and 140,820 tweets and uses a combination scoring strategy by gathering data from original source articles labeled as unreliable and then scoring and categorizing the content. Nevertheless, its outdated information and practice of labeling entire publishers as reliable or unreliable, as opposed to judging individual articles, restrict its overall functionality. Following on, MM-COVID [2] classifies articles as real or fake using fact-checking websites in a multilingual and multimodal combined data approach. This has the aim of including a wide variety of languages and media formats; it also faces some obstacles in fact-checking sources and does not give an exhaustive classification of visual indicators. Lastly, MMCoVaR looks to enhance the detection of misinformation by incorporating textual and visual information related to COVID vaccines. It uses a multimodal approach and utilizes machine learning for a more precise tertiary reliability classification. However, its focus on vaccines limits its general applicability to other domains of COVID misinformation [11].

In summary, the review of the existing body of research literature in this area presents both innovations and limitations in misinformation datasets. CoAID, ReCOVery, MM-COVID, and MMCoVaR have come up with substantial data but lack data completeness plus misinformation model accuracy, scope, and categorization without proper organized handling of image artifacts. These gaps give rise to the need for a more integrated approach towards collecting and utilizing multimodal data. This is where our CoVCues dataset plugs a hole by emphasizing image data and by creating novel classification for the gathered visual cues. It thereby adds a finer granularity for more accurate and effective misinformation detection. Table I displays a comparative analysis highlighting the difference between CoVCues and other relevant multimodal COVID datasets.

TABLE I. COMPARISON BETWEEN COVID DATASETS AND COVCUES

| *Dataset* | *CoAID* | *ReCOVery* | *MM-Covid* | *MM-CoVaR* | *CoVCues* |
|---|---|---|---|---|---|
| Textual Data | Yes | Yes | Yes | Yes | No |
| Image Data | No | No | Yes | No | No |
| VideoData | No | No | Yes | No | No |
| Binary Reliability Classification | Yes | Yes | Yes | No | Yes |
| Tertiary Reliability Classification | No | No | No | Yes | No |
| Textual Subcategories | No | No | No | No | No |
| Image Subcategories | No | No | No | No | Yes |

# III. COVCUES DATASET BUILDING PROCESS OVERVIEW

## A. Image Data Extraction and Cleaning

To retrieve the URLs from the existing COVID datasets, we used a series of Python scripts designed to extract the respective source URL fields using Python’s CSV library. The URLs were then written to separate text files based on whether they were considered reliable by the original datasets. Duplicates were removed using the “uniq” command. After some testing was done with the most common Python web scraping libraries, Scrapy was chosen for its ability to parse a very large number of websites both quickly and effectively. Beautiful Soup and Selenium were also considered, but they were not as efficient or user-friendly for working with many websites and for downloading images. This process needed two spiders: one for reliable URLs and one for unreliable URLs. The spiders were configured to use the URL text files as input, and download all images to their respective folders.

Due to the extremely large quantity of image data collected, it was necessary to cleanse the data to fit the project scope. Miscellaneous images with no impact on the research purpose such as icons, logos, and profile pictures needed to be removed for clarity. Additionally, outliers needed to be removed to ensure the data accurately portrayed the research subject. First, duplicate images were removed using a Python script to delete images with the same hash. This script covered both the reliable plus the unreliable folders and deleted approximately 2,500 images.

Second, images were sorted by size so favicon, icon, and emoji type images could be deleted. This helped identify icons, blurry images, and profile pictures. Small unrelated images were then easily removed. Lastly, the remaining pictures containing faces were sorted out using OpenCV [13], so that any leftover profile pictures could be manually identified and removed. Throughout this process, strange outlier images (like blank rectangles or images bearing no COVID connection) were reviewed and deleted at our discretion. This process was necessary for the data to be useful and more meaningful for training AI models later.

### B. *Image Data Categorization*

CoVCues images were put in reliable and unreliable categories based on source URLs. They were further classified into following subcategories: *signs, claims, fact-checks, ads,* and *miscellaneous* [2]. They were reorganized for clarity and to better fit the purpose. Figures 1 to 5 display examples from these subcategories. We utilized labels to categorize images based on patterns and infographics. Signs were images that signaled what the source was about and indicated if it could be considered reliable. Claims were images that could make claims, which could be fact or fake and had visuals that contained graphs/statistics. Fact-checks were images that marked statements or associated web contents as true or false. Ads were images portraying logos or products in a way to sell the user something. Lastly, miscellaneous ones were images unrelated to COVID that were not deleted during the data cleansing. The set of image artifacts were sorted into these sub-categories.

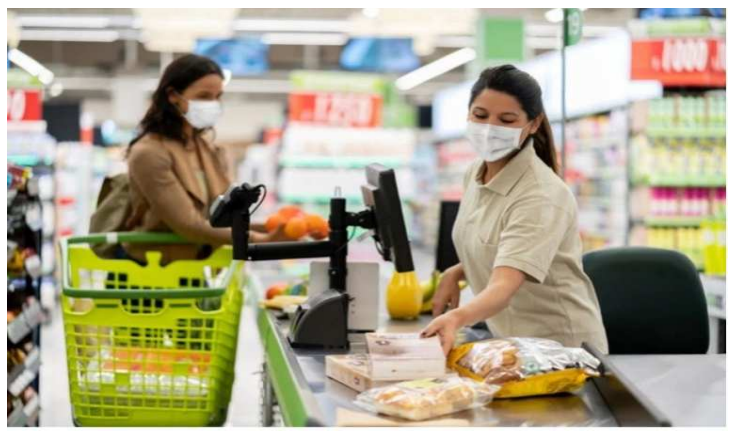

Fig. 1. A Sign Image Taken From Our CoVCues Dataset [15]

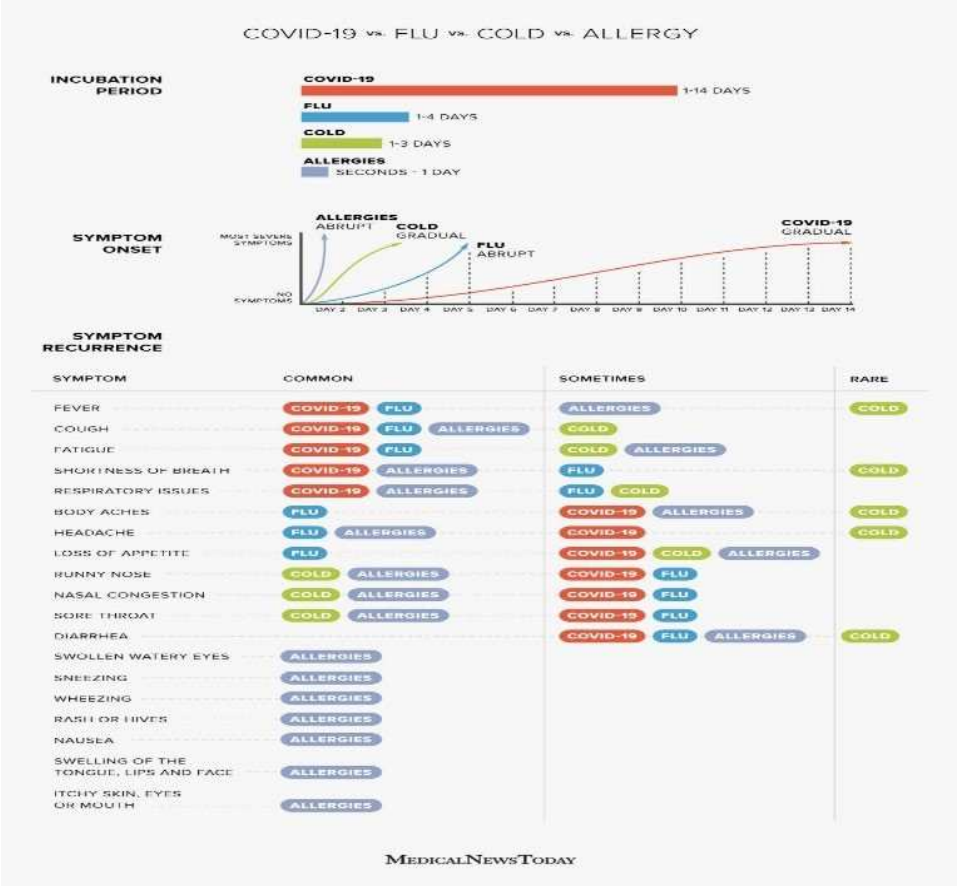

Fig. 2. A Claim Image Taken From Our CoVCues Dataset [15]

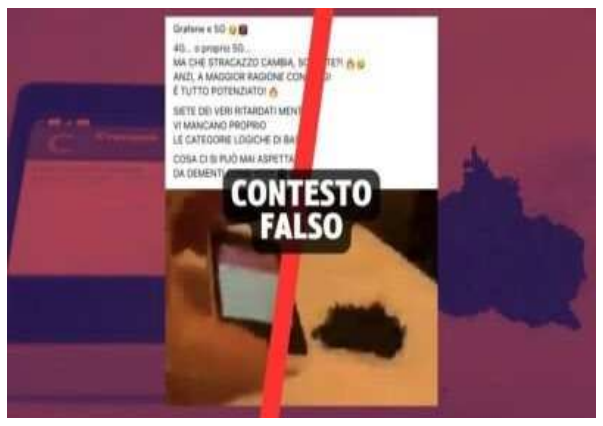

Fig.3. A Fact-Check Image Taken From Our CoVCues Dataset [15]

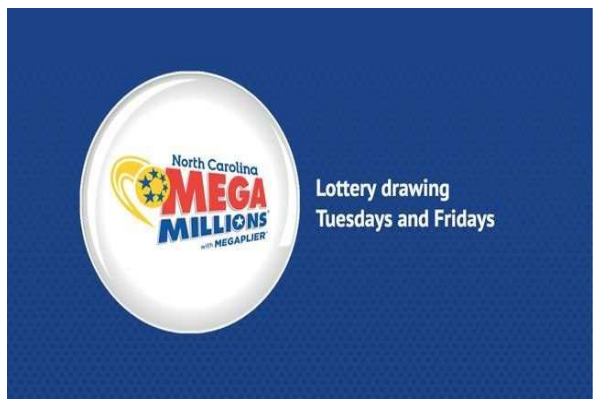

Fig. 4. An Advertisement Image Taken From Our CoVCues Dataset [15]

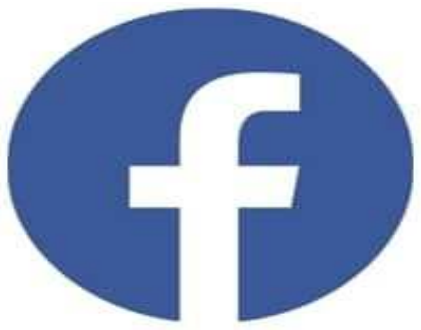

Fig. 5. A Miscellaneous Image Taken From Our CoVCues Dataset [15]

## IV. AI Models Used In CoVCues Analysis

After our CoVCues dataset images were cleaned up and sorted out, we applied the SVM (Support Vector Machine) model [14] to classify images into two distinct classes: reliable and unreliable. Each image was represented using both visual features and categorical metadata derived from the manual annotations. Each image's reliability labels were taken from the source URLs of the three referenced datasets (as discussed previously and as illustrated in Table II) and were labeled as either "reliable" or "unreliable" according to the labeling approach used in the source datasets. This image analysis approach involved a nuanced and proven AI technique. We chose SVM for the image classification process keeping in mind its effectiveness, hoping to dig deep into the image cues/patterns that stand out in terms of attributes linked to user perception of reliability. We considered the different subtleties in this regard.

TABLE II. URLs Collected From Other COVID Datasets Followed By URLs Used For CoVCues Dataset Building

| *Dataset* | *Reliable* | *Unreliable* | *Total* |
|---|---|---|---|
| CoAID | 4,253 | 371 | 4,624 |
| ReCoVery | 1,364 | 665 | 2,029 |
| MM-COVID | 11,256 | 3,151 | 14,407 |
| Total | 16,873 | 4,187 | 21,060 |
| *CoVCues Images* | *5,484* | *18,006* | *23,490* |

### A. SVM Model For CoVCues Image Analysis

The CoVCues image analysis process was based upon model classification into reliable and unreliable categories using an SVM algorithm. A support vector classifier was created with the help of GridSearchCV [14], and a model was constructed using the associated grid parameters. A split ratio was implemented to perform cross validation for model evaluation, given our limited resources (time, hardware, and memory) and due to the fact that cross validation is computationally expensive, as it trains the model multiple times (e.g., k-fold cross-validation trains the model k times), while a single train-test split is faster and more efficient. A split ratio of 70:30 was used to split the image data into training and testing samples for the model using the images from either reliable or unreliable URLs parsed. This image sampling for this process was done before the model was trained with both reliable and unreliable source images, and then categorized. We obtained a classification report out of this exercise to evaluate the model's performance and its accuracy. Table III shows this classification report depicting the model performance in terms of precision, recall, F1-score, and overall accuracy. The current accuracy of 69% indicates moderate reliability. It is to be noted and recognized that the model's predictive capacity was affected by the restricted image feature set and the binary labels derived from the parent dataset URLs.

TABLE III. CoVCues Analysis: Our SVM Model Accuracy

| ***Overall Accuracy:*** 0.69 | ***Precision*** | ***Recall*** | ***F1-Score*** | ***Support*** |
|---|---|---|---|---|
| Reliable | 0.65 | 0.80 | 0.72 | 40 |
| Unreliable | 0.74 | 0.57 | 0.65 | 40 |

## V. User Assessment Study Approach

A central focus of this paper is the user assessment study conducted to determine the perceived relevance, effectiveness, and potential impact of the CoVCues dataset [15] in supporting online COVID misinformation detection. In this study, participants were asked to self-identify as either OHI consumers or OHI providers, or both, and respond to a series of survey questions designed to assess how they perceive and utilize visual cues, textual data and both combined when determining the reliability of online COVID info contents.

Our survey included both quantitative and qualitative questionnaire items aimed at understanding participant perceptions of image reliability, the effectiveness of visual versus textual cues, and the combined influence of multimodal cues. We utilized these online survey questions to collect our user assessment data. Our questions were focused on gathering valuable insights into how different stakeholders interpret infographic content and respond to visual cues in health misinformation contexts. We also captured user perspectives on the role of prior exposure to misinformation and how it affects their assessment of visual cues. We did seek participants' take and thoughts on images portraying ground truth as reliability indicators by showing them a selected set of image artifacts from our CoVCues dataset and by asking them to rate their reliability or identify the real, factual visual cue or the fake, misinformation cue. Our hope is that this conducted survey and the research data obtained in the form of user responses will help set up the foundational groundwork for future multimodal misinformation related to further research and lead to better misinformation detection frameworks. In the process, our study assists to shed light on how visual cues can be used to increase accuracy, trust, and comprehension when processing OHI.

### A. Research Questions

Our primary objective of this study is to explore how the integration of visual cues alongside textual information influences users' ability to identify and assess COVID misinformation. More broadly speaking, we aim to gain insights into what both OHI consumers and OHI providers perceive as necessary for effectively mitigating the spread of false and misleading COVID information. The following are our main research questions addressed in this study:

Research Question 1: *How to validate the CoVCues dataset, i.e., how to determine its potential value and utility via a user study?*

Research Question 2: *How do stakeholders (content consumers versus providers) perceive reliability of visual, textual, and combined multimodal cues as part of online COVID info?*

Research Question 3: *How does prior exposure to misinformation impact user trust in visual cues and influence user ability to identify facts/truth versus fake/misinformation?*

Research Question 4: *To what extent do visual cues affect user perceptions about the trustworthiness of COVID information when compared to the impact of textual cues?*

## VI. Survey Data Collection Procedure

The survey questions were created using the Qualtrics tool. The participants for our research study included individuals from the academia, including faculty & students, and professionals/practitioners across different sectors from the industry, community partners organization (the non-academic community members). To better understand the participant pool, including their responses, and in an effort to organize the data into categories, we collected information on whether the participant was health info content consumers, content providers, or both. Participants were provided with a brief overview of the CoVCues project and were asked to determine the reliability of various visual cues (CoVCues images in pairs) and to provide feedback based on personal experience and judgement.

Academia made up a sizable portion of our survey participants, in addition to engineers, information tech and health professionals. In order to determine disparities in perception and information assessment practices, responses from academic and non-academic groups were examined. Survey questions included a mix of Likert scale ratings (ranging from 1 to 5) to capture the degree of perceived reliability and effectiveness of different cues, as well as binary (yes/no) questions for more straightforward feedback. Participants were presented with a series of questions in which they were asked to identify the more reliable images, in order to figure out the factual cues from the fake, misinfo ones. The images presented to participants were selected using a stratified random sampling process to ensure representation from both reliable and unreliable categories within CoVCues, and across each visual subcategory (sign, claim, fact-check, and advertisement). Each selected image pair included one reliable and one unreliable sample to balance exposure. This helped us explore how different types of online users perceive and determine image reliability. Open-ended questions were also included to gather qualitative feedback, critiques, and suggestions regarding the use of images in health misinfo detection. The results of this study, along with selected survey questions and the survey response data visualizations, are presented in the following sections. Figure 6 shows an example of the type of questions asked to survey participants.

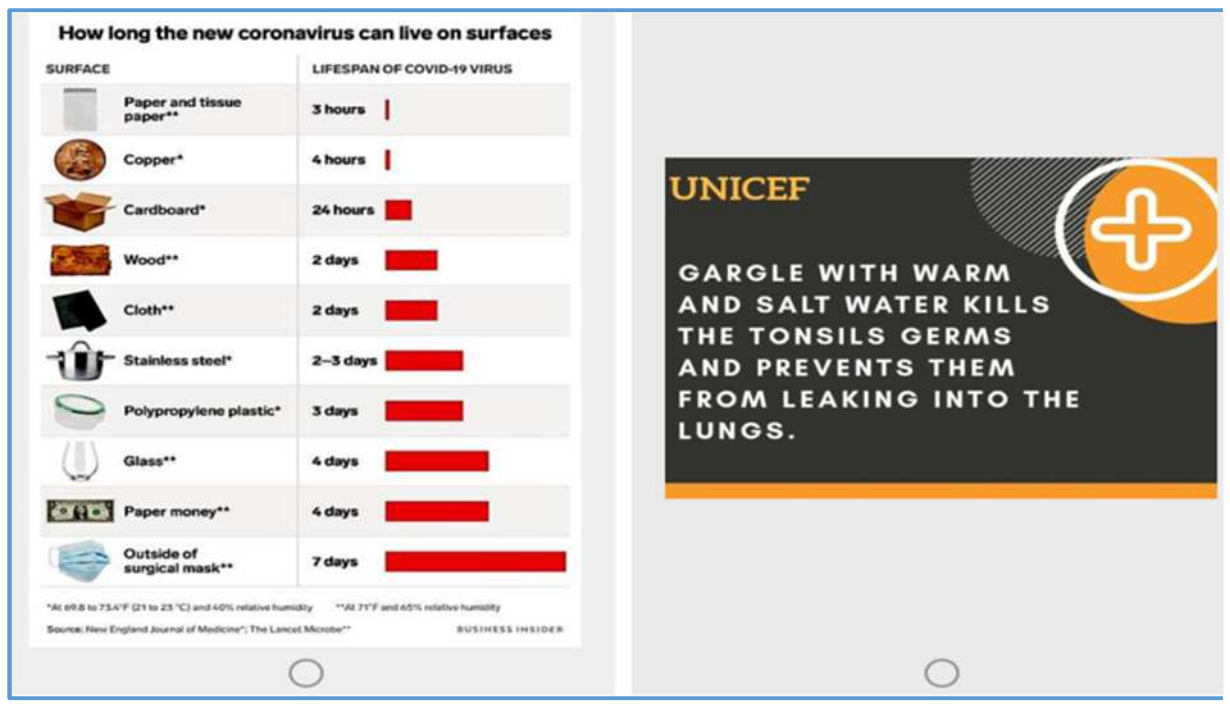

Fig. 6. An example of the type of questions in our online user survey.

## VII. User Assessment Study Results

A total of 78 participants, which included students, faculty and working professionals, completed our online survey. Out of these individuals, who formed our sample group of participants, 69 identified as online health info consumers, 5 identified as both content consumers and providers, and 4 identified as only content providers. Figure 7 summarizes this stakeholder distribution in percentage numbers based upon the composition of the survey participant group. In the following sections, we share details of the survey response data and report on the outcomes of our study.

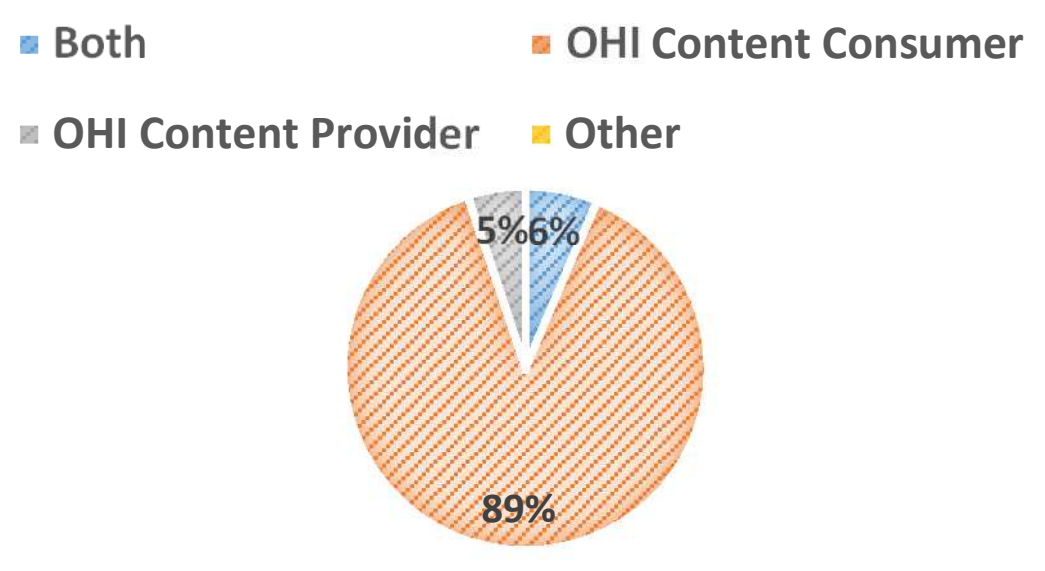

Fig. 7. Graphical summary data of responses to the following survey question: "*Select below the societal stakeholder category you fall under in context of this project i.e. when it comes to online health info (OHI).*" - Most respondents identified as content consumers, with few identifying as both consumer & provider, and fewer as only content provider.

### A. *Quantitative Data*

Our survey consisted of four (4) visual cue questions where for each pair of CoVCues images provided, participants were asked to select the image they think is the reliable or factual one. Furthermore, the participants were asked to provide their overall rating of the health misinfo topic/issue, including the importance of research, user awareness, and experience with online misinfo, and the effectiveness of visual cues in determining reliability. As already explained, Figure 7 represents one of the survey questions yielding quantitative data. We next discuss the survey questions that gather quantitative data on prior exposure to health misinfo. Figures 8 & 9 show this data, which indicates that 60 out of the 78 participants encountered COVID or other health related misinfo, while only 6 indicated otherwise. Notably, content consumers reported the highest frequency of misinformation exposure. Additionally, all participants, who identified as content providers, indicated prior exposure to health misinfo.

When analyzing participants' ability to distinguish between reliable (or factual) and unreliable (or misinfo) images through the received survey data, we observed that while the majority of responses were accurate, not all images were correctly identified by all participants. As illustrated in Figure 10, content providers demonstrated the highest consistency in correctly identifying reliable versus unreliable images. This finding suggests that prior knowledge or expertise may influence one's ability to accurately identify misinfo via visual cues, and that images alone may not always suffice in guiding user perceptions of reliability. Table IV shows the total number of reliable choices and unreliable choices by the participants in our survey. Participants were also asked to rate the perceived importance of different cue types - visual cues, textual cues, and a combination of both, so that we could determine the type of data tied to the reliability assessment of OHI. Additionally, we assessed which factors most strongly influence the user perception of credibility when processing online COVID info, as illustrated by Figure 11.

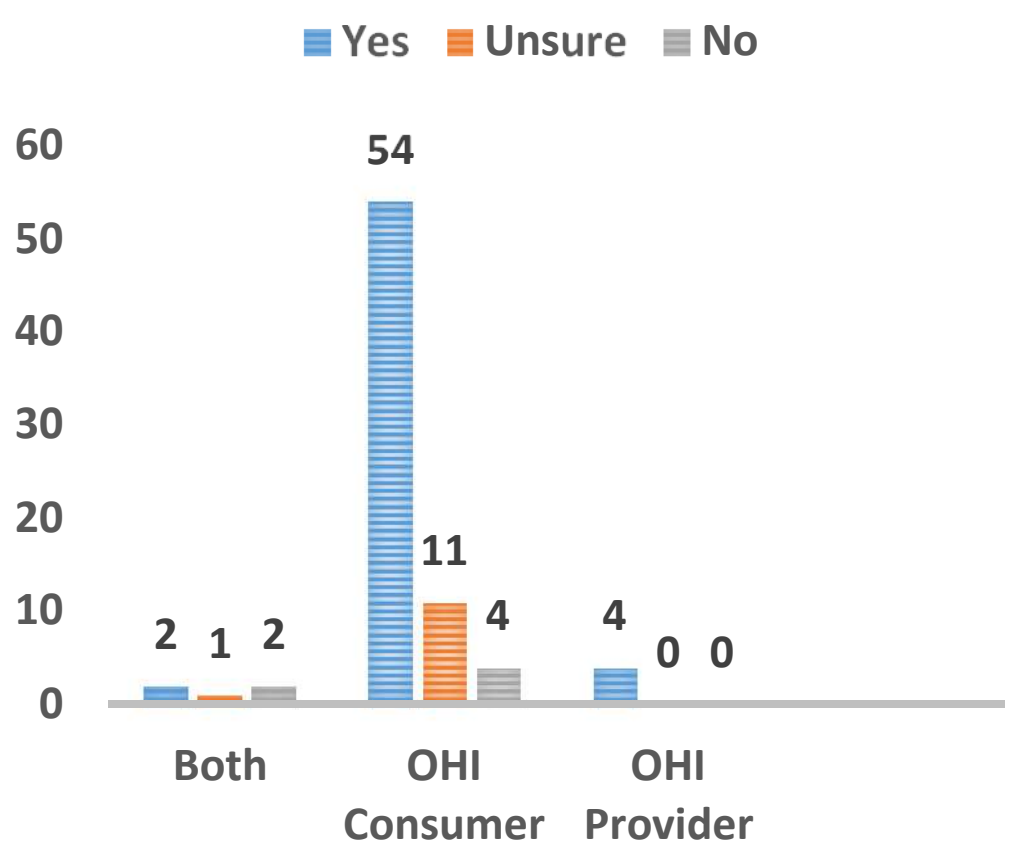

Fig. 8. Graphical bar chart summary data of responses to the following survey question: "*In your experience, have you encountered misinformation in relation to COVID, or other health related topics?*" - Most participants in all stakeholder categories reported encountering COVID or other health-related misinfo online.

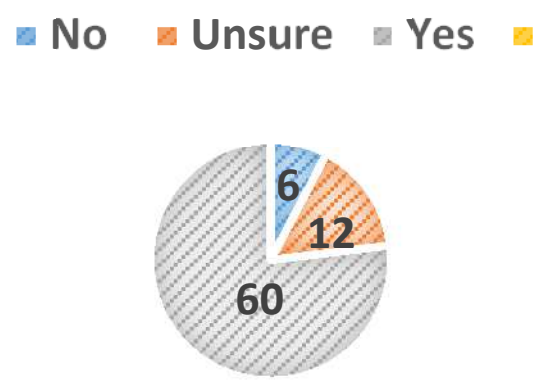

Fig. 9. Graphical pie chart summary data of responses to the following survey question: "*In your experience, have you encountered misinformation in relation to COVID, or other health related topics?*" - Most participants reported encountering COVID or health-related misinformation online.

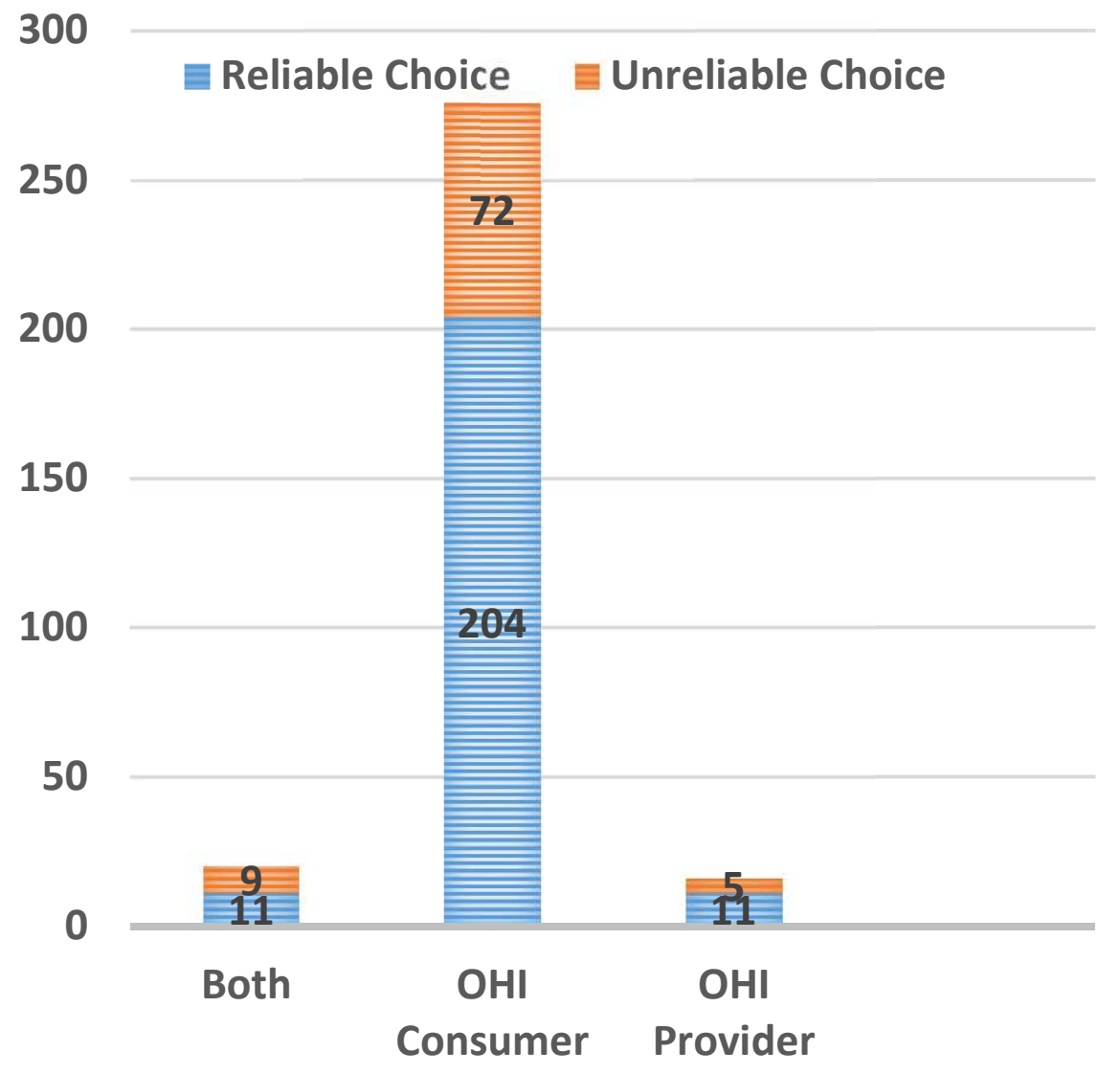

Fig. 10. Graphical summary data of overall participant responses, showing the total number of correctly identified images by all participants based on each stakeholder category and indicating that content providers demonstrated the highest consistency in correctly identifying reliable versus unreliable images.

TABLE IV. SUMMARY OF IMAGE CHOICES SHOWING USER VIEW POINTS BASED UPON SURVEY RESPONSES

| *Image Pairs In Survey Questions* | *Reliable Choice* | *Unreliable Choice* | *Proportion of Correct Choice* |
|---|---|---|---|
| Image Pair 1 | 65 | 13 | 83.3% |
| Image Pair 2 | 50 | 28 | 64.1% |
| Image Pair 3 | 61 | 17 | 78.2% |
| Image Pair 4 | 50 | 28 | 64.1% |

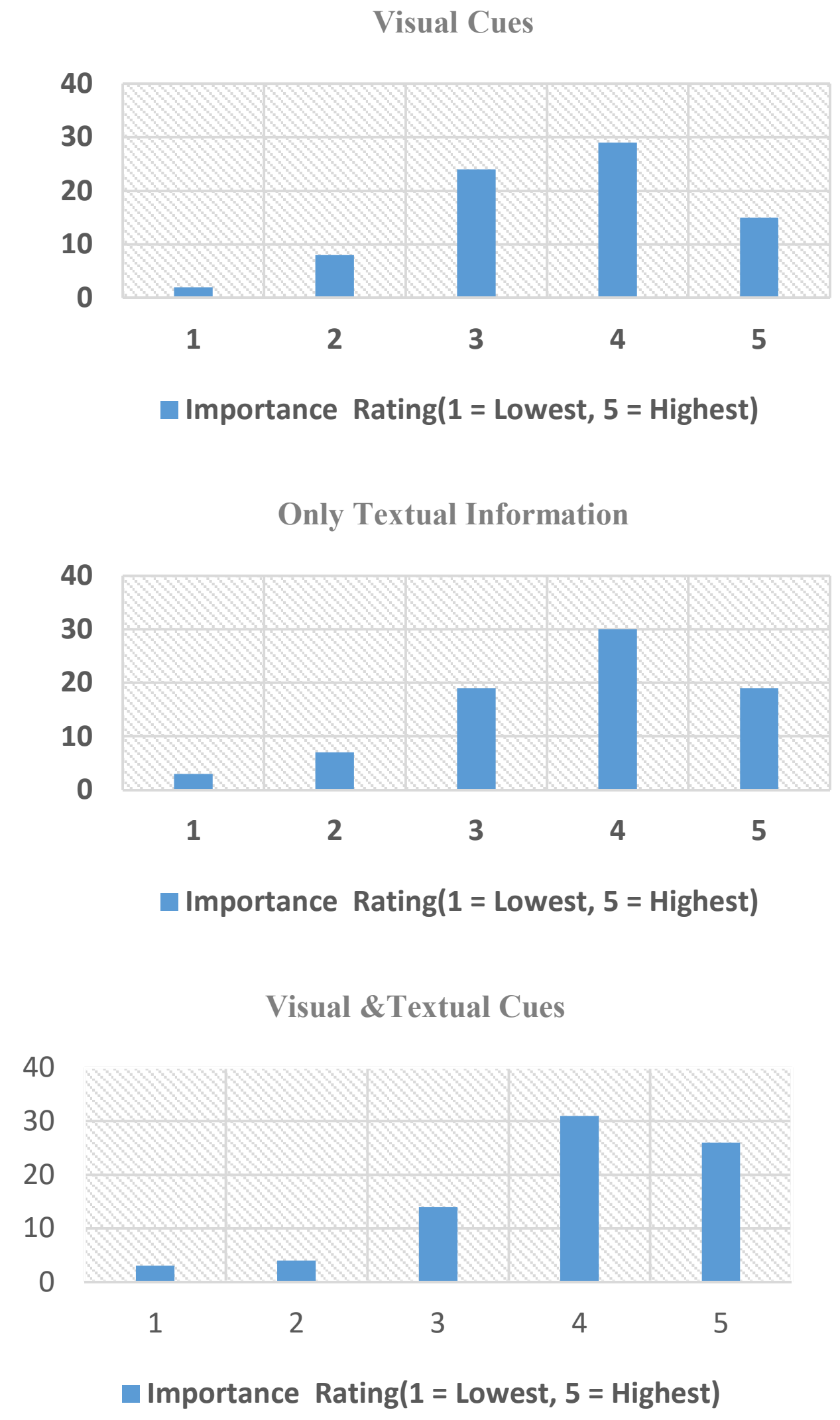

Fig. 11. Graphical summary data of responses to the following survey question: "*On a scale of 1 to 5, how effective are images in contributing towards the perception of reliability when used along with COVID related online posts and claims? Rate the utility value*" - Most participants rated the combination of visual cues and textual information as most important in identifying online health misinformation.

As seen in Figure 12, source credibility emerged as a key factor influencing users' trust in OHI. The two most significant factors influencing participants' take on COVID-related information were source credibility and the presence of images accompanied by supporting textual content.

## B. *Qualitative Data*

As part of the user assessment survey, participants were invited to provide open-ended feedback and suggestions regarding the CoVCues dataset and the overall survey questionnaire at the end of the survey. These responses we received formed the basis of our qualitative data and provided valuable insights into public perceptions of online health misinformation and the perceived utility of a multimodal dataset.

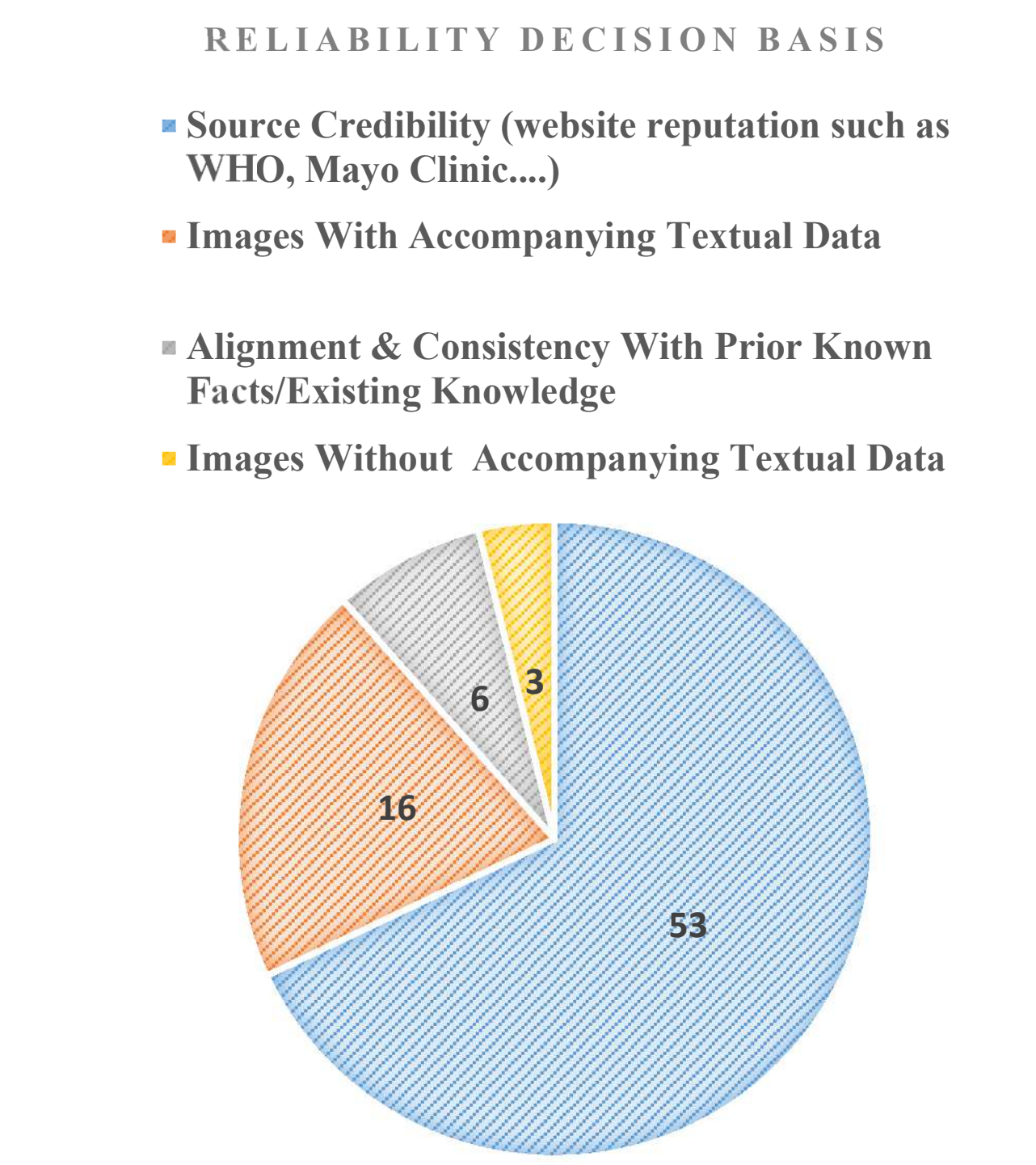

Fig. 12. Graphical summary data of responses to the following survey question: "*On a scale of 1 to 5, how effective are images in contributing towards the perception of reliability when used along with COVID related online posts and claims? Rate the utility value*"- Most participants rated the combination of visual cues and textual info as most important in identifying online health misinformation.

Figure 13 represents a summarized representation of our overall qualitative data collected through our survey responses in the form of a word cloud diagram that contains the keywords and/or most common words, as used by the study participants. An overall analysis of this received data shows a trend of certain focus keywords emerging from the participant responses when asked about their feedback regarding the CoVCues dataset. This word chart, as displayed in Figure 13, was developed to capture these keywords and to show the magnitude of their implications.

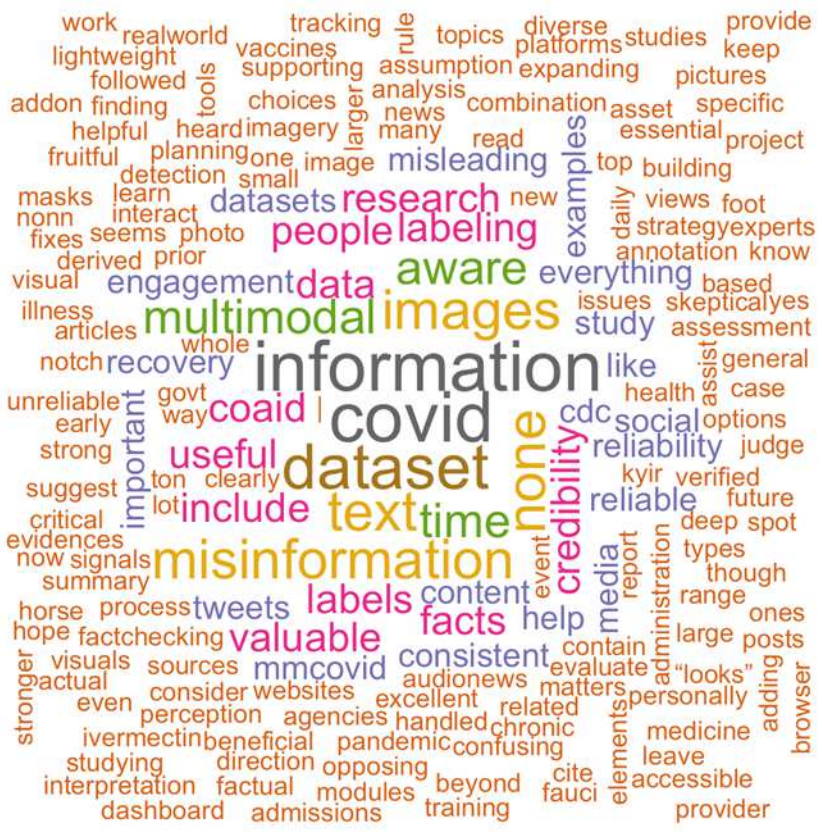

Fig. 13. A word cloud diagram highlighting the keywords and/or most common terms used in the responses to the ending survey question (seeking overall feedback comments, suggestions, and inputs) that we received from the participants in our user study.

A recurring theme among these received responses was the difficulty participants faced in discerning trustworthy information during the COVID infodemic. Many shared that, although they regularly encountered misinformation online, they often struggled to identify reliable sources. Several participants expressed strong support for an extensive multimodal dataset, such as CoVCues, that includes both textual and visual cues. They emphasized that such a dataset could aid both content consumers and content providers in verifying plus disseminating accurate information. However, some respondents also voiced skepticism, sharing that due to the overwhelming amount of false information plus the growing spread of misinformation, many users have grown hesitant to trust any online content regardless of its source or presentation.

One participant explained: *"For any issue, such as COVID-19, it is critical to evaluate information and media that both support and challenge the prevailing narrative, especially when backed by evidence and factual reasoning."* Another participant emphasized the importance of clarity in design and structure, stating: *"I recommend implementing a consistent and rigorous labeling and annotation strategy from the start to classifying content as 'reliable,' 'unreliable,' or 'misleading' as this will significantly enhance the dataset's usefulness."*

Concerns about broader public skepticism were also raised by a few participants, as evident in the following comment: *"I think the way the COVID pandemic was handled has made many people very skeptical of government agencies, which makes them question or ignore most health-related information."* From a usability standpoint, one respondent suggested improvements for accessibility as seen in the following feedback: *"Larger image sizes would be helpful, especially for users who have difficulty reading small text."*

Overall, the feedback comments and suggestions reflected a mix of optimism and caution. While there was clear appreciation for the effort to create CoVCues as a structured, multimodal COVID dataset resource, the responses also underscored the importance of transparency, thoughtful design, and trust-building when developing resources or data to combat the still existent infodemic challenge due to the presence of COVID misinformation in the health domain, including the online space.

## VIII. Statistical Analysis Of Survey Data

To assess the reliability and the perceived effectiveness of multimodal data, including both visual and textual cues, we built the CoVCues dataset, and performed a preliminary user study involving 78 participants. We computed descriptive statistics and 95% confidence intervals (CI) for participants' accuracy in distinguishing between reliable and unreliable images, as well as their perceived usefulness ratings for visual, textual, and combined cues. All ratings were reported on a 5-point Likert scale, with 1 indicating "not useful at all" and 5 indicating "extremely useful." The average number of correctly identified images across all participants was 2.9 out of 4 (95% Confidence Interval: 2.72-3.08), indicating a moderate level of accuracy in distinguishing between reliable and unreliable COVID visual cues. This suggests that while many participants were able to make informed judgments based on these cues, there is still room for improvement in how such cues are designed and presented.

Regarding the perceived usefulness of different information types, the combined use of textual and visual cues received the highest average rating of 3.6 on a 5-point Likert scale (95% CI: 3.6 ± 0.22). This result highlights a strong user preference for multimodal content when evaluating the reliability of OHI, reinforcing the foundational motivation for the CoVCues dataset. To assess the adequacy of our sample size, a post-hoc power analysis was conducted based on an observed effect size (Cohen's f = 0.239) derived from the comparison of visual cue ratings across misinformation exposure groups (Yes, No and Unsure). With three groups and 26 participants per group (n = 39), the resulting power was 0.42. This suggests that the current study was underpowered to detect effects of this size and highlights the need for a larger sample in future work to improve the robustness and generalizability of the findings. This finding has made us realize that a larger and more diverse sample in future studies would enhance statistical robustness and improve the generalizability of our present findings.

## IX. Concluding Summary

Our main contribution in this paper is the unique, maiden user assessment based research study conducted with the new CoVCues dataset in an effort to determine the potential of using multimodal cues, including visual cues, for helping with COVID misinformation detection research. In the process, we perform a novel analysis, including a first-time sub-categorization and granular level organization of the CoVCues dataset. CoVCues is an exclusive image artifacts collection to help extend the value of the multimodal cues, especially infographic image contents, in order to aid COVID infodemic researchers. Hence, we envision that this study could prospectively pave the path for new research and development in this emerging area, as it showcases how visual cues can increasingly play a pivotal role in online health misinformation recognition, facilitating the development of more effective and stronger COVID misinformation detection mechanisms.

Additionally, the research data obtained through our study will assist researchers in understanding user perceptions around some newly explored aspects of image reliability (in the context of COVID misinformation), the effectiveness of visual versus textual cues, and the combined effect of multimodal cues on the ability of content consumers & providers to identify misinformation, as well as distinguish between facts and fakes. We have been able to draw valuable insights on how different stakeholders process image contents and respond to visual cues in the context of health misinformation. We have also been able to throw light on how previous exposure to misinformation can add value to user perceptions and abilities by increasing accuracy, confidence, and conviction when dealing with OHI. Moreover, our obtained survey data clearly indicates that health misinformation continues to be a real potent problem in today's society with OHI content consumers and OHI content providers both being susceptible to it. Also, our study indicates that we are not out of the woods when it comes to the COVID infodemic problem even post pandemic. We next discuss our shortcomings in this work.

### *A. Challenges and Limitations*

We did face some challenges while building the CoVCues dataset. During the initial extraction attempts for gathering the visual cues (image artifacts) using Beautiful Soup, we encountered frequent time outs due to the large volume of data being processed. Consequently, the Scrapy framework was adopted, and this helped us to overcome these issues plus do improved handling of large-scale web scraping. However, certain websites still managed to block the scraper despite our bypass configurations. Although we were able to successfully collect thousands of images, additional challenges emerged during the post-processing phase, particularly in removing duplicate and blurry images. The automated scripts developed for this task could not perform the necessary review to ensure data quality, and we ended up doing that manually ourselves. Notably, this was quite a time consuming task.

Manual intervention was also required at multiple stages of data cleansing, such as the removal of duplicate or blurry images and the assignment of subcategories to the image artifacts. These manual processes introduce potential subjectivity and bias, which may impact the accuracy and generalizability of our dataset. Challenges also arose during AI model development for our image analysis part. While using Keras to train classification models that differentiated between reliable and unreliable images, frequent issues related to overfitting and system crashes were encountered. The dataset size, consisting of approximately 20,000 images, led to frequent memory overuse and runtime errors during this AI model training. These obstacles required iterative adjustments to our model design and system resources to ensure stability and effective learning.

As an initial user assessment and ongoing research effort, this work is subject to some limitations stemming from both the source datasets (MM-COVID, CoAID, and ReCOVery) and the nature of data processing involved. Firstly, the CoVCues dataset assigns reliability labels based on the parent URLs' ratings from the original datasets. This approach does not fully account for the possibility that reliable sources may sometimes include misleading visuals, or vice versa, leading to potential mislabeling at the source dataset image level and present annotations being used. Moreover, the rapidly evolving nature of COVID info introduces temporal limitations. As scientific understanding of the virus has advanced, what was previously labeled as misinformation may no longer be valid, and vice versa. The current dataset only captures misinformation definitions as they were understood at the time of each source dataset's publication.

### B. *Future Scope Of Work*

There are several promising directions for expanding this research and building on this preliminary research study. First, the CoVCues dataset itself remains a work-in-progress under active development, with ongoing data cleansing, subcategorization, and quality enhancement efforts. Improvements in these aspects will allow for more robust training of AI models and enable deeper analysis with the aid of our collected multimodal cues, including visual cues, to result in more robust COVID misinformation detection down the line.

The current user assessment study phase serves as an initial validation of the dataset's potential utility. Through a structured Qualtrics survey, our study participants were presented with a subset of CoVCues images to determine their ability to differentiate between reliable and unreliable visuals. Based on these findings, future efforts will include refining the survey instrument, continuing with the participant recruitment process, plus the collection of more user data, and addressing potential ambiguities or image quality issues in the CoVCues dataset.

Additionally, future research will focus on performing further examinations using other evaluation techniques (non-user assessment studies) to test how incorporating visual cues (from CoVCues) alongside textual data can work successfully for improving user trust, comprehension, and strengthening overall model performance in future misinformation detection tasks. The expanded CoVCues dataset, which we have plans to work on, would also include new multimodal components, such as video and audio, which have not been explored in the COVID context, in order to enhance its applicability across broader misinformation scenarios. Incorporating tertiary classification frameworks, inspired by approaches such as MM-CoVaR, may further strengthen the CoVCue dataset's accuracy and utility.

Further down later, we plan to test the enriched and enhanced CoVCues dataset through larger and more comprehensive usability studies that simulate real-world misinformation detection scenarios. Insights gained from these studies, coupled with more diverse user feedback, will result in appropriate refinement, fine tunings, and new additions to the broader development of the future expanded version of CoVCues.